# Determining the acoustoelastic effect of longitudinal waves propagating inclined to principal stress directions in concrete: theory and experimental validation

Hao Cheng[a,*], Katrin Löer[b], Max A.N. Hendriks[a,c], Yuguang Yang[a]
[a] Department of Engineering Structures, Faculty of Civil Engineering and Geosciences, Delft University of Technology, 2628 CN Delft, the Netherlands
[b] Department of Geoscience and Engineering, Faculty of Civil Engineering and Geosciences, Delft University of Technology, 2628 CN Delft, the Netherlands
[c] Department of Structural Engineering, Faculty of Engineering, Norwegian University of Science and Technology, 7491 Trondheim, Norway
[*] Corresponding author: h.cheng-2@tudelft.nl

**Abstract**
The concept of acoustoelasticity pertains to changes in elastic wave velocity within a medium when subjected to initial stresses. However, existing acoustoelastic expressions are predominantly developed for waves propagating parallel or perpendicular to the principal stress directions, where no shear stresses are involved. In our previous publication, we demonstrated that the impact of shear stresses on longitudinal wave velocity in concrete, when body waves propagate in the shear deformation plane, is negligible. This finding allows us to revise the acoustoelastic expression for longitudinal waves propagating inclined to the principal stress directions in stressed concrete. The revised expression reveals that the acoustoelastic effect for such longitudinal waves can be expressed using acoustoelastic parameters derived from waves propagating parallel and perpendicular to the uniaxial principal stress direction. To validate our theoretical statement, experiments were conducted on a concrete cylinder subjected to uniaxial stress. Despite slight fluctuations in the experimental observations, the overall trend of acoustoelastic effects for inclined propagating longitudinal waves aligns with the theory. This proposed theory holds potential for monitoring changes in the magnitudes and directions of principal stresses in the plane stress state.

## 1. Introduction

Acoustoelasticity is a phenomenon that describes how an elastic wave changes in an initially-stressed medium, including variations in velocity and polarization direction. The study of acoustoelasticity can be traced back to the 1820s [1], and in the latter half of the 20th century, the theoretical framework for acoustoelasticity was established by incorporating nonlinear elasticity into the equation of motion [2-4]. This framework has been applied to analyze changes in both bulk waves [5, 6] and surface waves [7, 8]. However, the existing expressions of acoustoelasticity are predominantly limited to cases where wave propagation is parallel or perpendicular to the axes of principal deformations. However, in concrete infrastructure monitoring, elastic waves do not always propagate along predetermined directions. For example, in monitoring a concrete bridge deck under moving vehicle loads, the stress conditions change over time. In such cases, it is impossible to consistently align wave propagation with the principal deformation directions.

When there is an angle between the body wave propagation direction and the principal stress direction, a new coordinate system can be established parallel and perpendicular to the wave propagation direction. By computing the stress matrix within this new coordinate system, shear stresses can be identified. The presence of shear stresses indicates the existence of shear deformations, which are characterized as isochoric plane deformations [9]. In such deformations, a set of line elements with a given reference orientation does not change in length or orientation. Some efforts in the literature aim to address the acoustoelastic effect of waves propagating at an angle to the principal stress directions in metals [10-14]. However, a critical assumption in these studies is that the influence of shear stresses on body wave velocity is negligible, which has not been proven valid in the aforementioned articles. In our previous work [15, 16], we incorporated normal and shear deformations in the governing equation of acoustoelasticity. Our numerical examples using concrete parameters demonstrated that while shear strains significantly affect transverse wave velocity, they have a minimal impact on longitudinal wave velocity.

In this paper, building on the main conclusions from our previous work, we aim to quantify the acoustoelastic effect of longitudinal waves propagating inclined to the principal deformations in concrete through both theoretical and experimental investigations. We begin with a brief overview of current efforts to quantify the acoustoelastic effect in concrete in Section 2, highlighting the significance of our research. Section 3 presents the theoretical background for determining the acoustoelastic effect of longitudinal waves propagating inclined to principal deformations. To validate our proposed theory, we design a set of experiments using a concrete cylinder as the specimen, detailed in Section 4 along with the experimental setup, sensor layout, and signal processing techniques. Section 5 presents the acoustoelastic effects of longitudinal waves propagating inclined to the principal stress directions acquired from experiments. To further validate these effects, they are used to back-calculate the acoustoelastic parameters for longitudinal waves propagating parallel and perpendicular to the uniaxial principal stress direction. These back-calculated parameters are then compared with values reported in the literature, as detailed in Section 6. An discussion is provided in Section 7, focusing on potential sources of error and potential applications of the proposed theory.

## 2. Brief review on the studies and applications of acoustoelasticity in concrete

Unlike previous research focused on the acoustoelastic effect in polymers [2], metallic materials [17-19], and woods [20-22], the studies and applications of acoustoelasticity in concrete has emerged relatively recently. Most existing studies have aimed to establish empirical relationships between stress and wave properties, such as wave velocity [23-27] and coherency [28], to quantify the changes in the velocity-stress relationship in concrete. These studies often use wave interferometry, which compares waveforms and measures velocity changes before and after a disturbance [29] to detect stress-induced velocity changes. Compared with the arrival time picked-based ultrasonic pulse velocity (UPV) methods, such as Akaike information criterion-based UPV [30], wave interferometry is much more sensitive to weak changes in the medium [31, 32].

Wave interferometry can be applied to both ballistic [33] and coda wave components [34-37]. However, when used in the coda component, which is the tail of the waveform [38], the observed acoustoelastic effect represents an ensemble average of the effects on both longitudinal and transverse waves. Consequently, caution is needed when interpreting results from studies that attempt to extract acoustoelastic parameters of pure longitudinal and transverse waves from the coda [39-41]. Due to the difficulty of quantitatively assessing the acoustoelastic effect in concrete, inferring the stress field from the wave velocity field using coda components remains infeasible [34-37].

Despite the challenges, researchers have continued to investigate the acoustoelastic effect in concrete. Nogueira and Rens [42] conducted tests on 13 concrete specimens with 9 different mixtures to obtain third-order elastic constants, known as Murnaghan constants [43], which are closely related to the acoustoelastic response of a material. They used arrival time picker-based UPV to estimate velocity, revealing significant fluctuations in these constants across specimens, even for those with the same mixture. Lillamand et al. [44] used wave interferometry on the ballistic wave component to determine five acoustoelastic parameters for longitudinal and transverse waves propagating parallel or perpendicular to the uniaxial principal stress direction. These parameters consist of Lamé parameters and Murnaghan constants, which will be further introduced in Section 3. Zhong et al. [33] sought to assess the stress level of a real structure using laboratory-calibrated acoustoelastic parameters and adopted wave interferometry in the signal processing of both lab and in-situ measurements. Please note that the aforementioned studies relied on a single transmitter-receiver pair to derive acoustoelastic parameters or Murnaghan constants and focused on scenarios where the wave propagation direction aligns with one of the principal stress directions.

Based on this brief review, three observations can be made. First, wave interferometry is more effective than UPV in detecting stress-induced velocity changes in concrete due to its heightened sensitivity. Second, significant fluctuations in Murnaghan constants reported by Nogueira and Rens [42] indicates that relying on a single transmitter-receiver pair to derive Murnaghan constants or acoustoelastic parameters for a sample may be risky due to the spatial variation of mechanical properties in concrete [45]. Third, current publications on concrete focus on scenarios where the wave propagation direction aligns with one of the principal stress directions, with no studies addressing waves traveling at an angle to principal stress directions.

## 3. Theoretical background

In this section, we will revise the theory of acoustoelasticity specifically for longitudinal waves in a plane stress state for the ease of our reader. The plane stress state is commonly encountered in concrete structures such as girders during service. Our analysis assumes that the medium is isotropic, compressible, and elastic.

It should be noted that although concrete is not a homogeneous material, the quasi-uniform distribution of coarse aggregates makes the velocity of longitudinal and transverse waves in unloaded concrete nearly isotropic when the wavelength is much smaller than the sample geometry and the measurements are taken over sufficiently large length scales. In this scenario, the body wave will probe the concrete as an isotropic effective medium, where the probed mechanical properties are an average of the different phases [46], such as mortar and coarse aggregates. Additionally, after traveling a sufficient distance, the body waves are in the far-field regime and can be assumed to be plane waves [47]. The theory in this chapter is based on the theoretical framework of acoustoelasticity. More details regarding this theoretical framework can be found in our previous work [15, 16].

This section is organized in the following manner. Section 3.1 presents the general expression of acoustoelasticity for longitudinal waves. Subsequently, in Section 3.2, we focus on the acoustoelasticity for longitudinal waves propagating along one of the principal stress directions in the plane stress state. In Section 3.3, the acoustoelastic expression for longitudinal waves propagating inclined to the principal stress directions in the plane stress state is given.

### 3.1 Theoretical background of acoustoelasticity for longitudinal waves

Assuming a longitudinal wave propagates along one of the three principal stress directions, as shown in Figure 1, we establish a coordinate system with the wave propagation direction as the $x$-axis and the other principal stress directions as the $y$- and $z$-axis. The principal stress applied along the $x$-axis is denoted as $\sigma_1$, while the principal stresses along the $y$- and $z$-axes are denoted as $\sigma_2$ and $\sigma_3$, respectively. The principal strains induced by the principal stresses are represented as $e_1$, $e_2$ and $e_3$, which are along $x$-, $y$-, and $z$-axes, respectively. The relation between longitudinal wave velocity and principal strains in the scenario shown in Figure 1 can be expressed as [5]:

$$\rho v^2 = C_{11} + (3C_{11} + C_{111})e_1 + (C_{12} + C_{112})(e_2 + e_3) , \qquad (1)$$

where $\rho$ denotes the density of the material when no load is applied, and $v$ represents the velocity of longitudinal wave propagating in the direction shown in Figure 1. The coefficients $C_{11}$ and $C_{12}$ are second-order elastic coefficients that depend on the Lamé parameters, $\lambda$ and $\mu$. For an isotropic material, the values of the second-order elastic coefficients can be found in Table 1. The coefficients $C_{111}$ and $C_{112}$ are third-order elastic coefficients that depend on the Murnaghan constants $l$, $m$, and $n$ [43]. These third-order elastic coefficients exhibit the following symmetry in isotropic materials [48]:

$$C_{IJK} = C_{IKJ} = C_{JIK} = C_{JKI} = C_{KIJ} = C_{KJI} , \qquad (2)$$

where $I, J, K \in \{1, 2, 3\}$. For an isotropic material, the values of the third-order elastic coefficients can be found in Table 2. The derivation process of Equation (1) can be found in Appendix A.

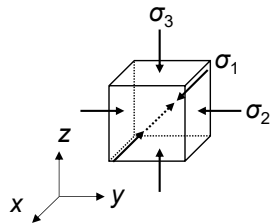

Figure 1. Diagram depicting principal stress directions and wave propagation direction within the coordinate system (solid arrow: principal stress directions; dotted arrow: longitudinal wave propagation direction).

Table 1. Second-order elastic coefficients for isotropic materials.

| Second-order elastic coefficients | Representation using Lamé parameters |
|---|---|
| $C_{11}=C_{22}=C_{33}$ | $\lambda+2\mu$ |
| $C_{12}=C_{13}=C_{21}=C_{23}=C_{31}=C_{32}$ | $\lambda$ |
| $C_{44}=C_{55}=C_{66}$ | $\mu$ |
| All others | 0 |

Table 2. Third-order elastic coefficients for isotropic materials.

| Third-order elastic coefficients | Representation using Murnaghan constants |
|---|---|
| $C_{111}=C_{222}=C_{333}$ | $2l+4m$ |
| $C_{112}=C_{113}=C_{221}=C_{223}=C_{331}=C_{332}$ | $2l$ |
| $C_{123}$ | $2l-2m+n$ |
| $C_{441}=C_{552}=C_{663}$ | $m-n/2$ |
| $C_{442}=C_{443}=C_{551}=C_{553}=C_{661}=C_{662}$ | $m$ |
| $C_{456}$ | $n/4$ |
| All others | 0 |

It is important to highlight that Equation (1) is derived in the natural frame of acoustoelasticity. This option is grounded in the fact that the natural velocity is inversely proportional to the time of travel, eliminating the need to correct the length of wave path using strain measurements [5].

### 3.2 Acoustoelasticity for longitudinal waves propagating along one of the principal stress directions in the plane stress state

Assuming that a uniaxial stress, $\sigma_1$, is applied along the *x*-axis (see Figure 1), and considering the propagation of a longitudinal wave along the same axis. In this case, the velocity of the longitudinal wave can be expressed as follows:

$$\rho v^2 = \lambda + 2\mu + (3\lambda + 6\mu + 2l + 4m)e_1 + (\lambda + 2l)(e_2 + e_3)$$
$$= \lambda + 2\mu + (3\lambda + 6\mu + 2l + 4m)\frac{\sigma_1}{E} - 2(\lambda + 2l)\frac{\upsilon \sigma_1}{E} \quad , \quad (3)$$

where $\upsilon$ represents the static Poisson ratio and *E* is the static elastic modulus. Although Lamé's parameters are known as linked to the elastic modulus and Poisson ratio, it is important to note that in viscoelastic materials like concrete, the static Poisson ratio and static elastic modulus cannot be represented using Lamé parameters in Equation (3). This is because these Lamé parameters are derived from the constitutive equation for high-frequency ultrasonic waves and may not adequately describe the constitutive relation of viscoelastic materials under static loading conditions, where the strain rate is much lower than that of ultrasonic waves. Equation (3) can be further simplified into the following form by introducing the acoustoelastic parameter $A_1$:

$$v = v^* \sqrt{1 + A_1 \sigma_1} \quad , \quad (4a)$$

where:

$$A_1 := \frac{3\lambda + 6\mu + 2l + 4m - 2\upsilon(\lambda + 2l)}{E(\lambda + 2\mu)} \quad , \quad (4b)$$

The longitudinal wave velocity without external load applied is represented as $v^*$, which is equal to $[(\lambda+2\mu)/\rho]^{1/2}$.

When the longitudinal wave propagates along the *x*-axis while the uniaxial stress, $\sigma_2$, is applied along the *y*-axis, the longitudinal wave velocity is:

$$\rho v^2 = \lambda + 2\mu + (3\lambda + 6\mu + 2l + 4m)e_1 + (\lambda + 2l)(e_2 + e_3)$$
$$= \lambda + 2\mu - (3\lambda + 6\mu + 2l + 4m)\frac{\upsilon \sigma_2}{E} + (\lambda + 2l)\left(\frac{\sigma_2}{E} - \frac{\upsilon \sigma_2}{E}\right) \quad , \quad (5)$$

Similarly, Equation (5) can be simplified by introducing the acoustoelastic parameter $A_2$:

$$v = v^* \sqrt{1 + A_2 \sigma_2} \quad , \quad (6a)$$

where:

$$A_2 := \frac{\lambda + 2l - \upsilon(4\lambda + 6\mu + 4l + 4m)}{E(\lambda + 2\mu)} \quad , \quad (6b)$$

When the longitudinal wave propagates along the *x*-axis and the bi-axial stresses are applied along the *x*- and *y*-axis, the longitudinal wave velocity in this plane stress state is:

$$\rho v^2 = C_{11} + (3C_{11} + C_{111})e_1 + (C_{12} + C_{112})(e_2 + e_3)$$

$$= \lambda + 2\mu + (3\lambda + 6\mu + 2l + 4m)\left(\frac{\sigma_1}{E} - \frac{\upsilon\sigma_2}{E}\right) + (\lambda + 2l)\left(\frac{\sigma_2}{E} - \frac{\upsilon\sigma_1}{E} - \frac{\upsilon\sigma_1}{E} - \frac{\upsilon\sigma_2}{E}\right)$$

$$= \lambda + 2\mu + (3\lambda + 6\mu + 2l + 4m)\frac{\sigma_1}{E} - 2(\lambda + 2l)\frac{\upsilon\sigma_1}{E}$$

$$- (3\lambda + 6\mu + 2l + 4m)\frac{\upsilon\sigma_2}{E} + (\lambda + 2l)\left(\frac{\sigma_2}{E} - \frac{\upsilon\sigma_2}{E}\right),$$

(7)

where $\sigma_1$ and $\sigma_2$ represent the principal stresses along the *x*- and *y*-axis, respectively. Note that Equation (7) is the combination of Equation (3) and Equation (5) and can be simplified as:

$$v = v^* \sqrt{1 + A_1\sigma_1 + A_2\sigma_2} .$$

(8)

### 3.3 Acoustoelasticity for longitudinal wave propagating inclined to principal stress directions in the plane stress state

In a plane stress state, we assume that the principal stresses are applied along the *x*- and *y*-axes, as illustrated in Figure 2(a), and the stress in *z* axis is zero. The propagation direction of a longitudinal wave is indicated by the dotted arrow. To make the acoustoelasticity applicable in this scenario, we need to align the axes of the coordinate system parallel and perpendicular to the wave propagation direction, as depicted in Figure 2(b). The normal and shear stresses in the new rotated coordinate system, denoted as *x'* and *y'*, can be determined by:

$$\begin{bmatrix} \sigma_{11} & \sigma_{12} \\ \sigma_{21} & \sigma_{22} \end{bmatrix} = \begin{bmatrix} \cos\theta & \sin\theta \\ -\sin\theta & \cos\theta \end{bmatrix} \begin{bmatrix} \sigma_1 & 0 \\ 0 & \sigma_2 \end{bmatrix} \begin{bmatrix} \cos\theta & -\sin\theta \\ \sin\theta & \cos\theta \end{bmatrix}$$

$$= \begin{bmatrix} \cos^2(\theta)\sigma_1 + \sin^2(\theta)\sigma_2 & \sin(\theta)\cos(\theta)(\sigma_2 - \sigma_1) \\ \sin(\theta)\cos(\theta)(\sigma_2 - \sigma_1) & \sin^2(\theta)\sigma_1 + \cos^2(\theta)\sigma_2 \end{bmatrix},$$

(9)

where $\theta$ is the inclination of the wave propagation, and $\sigma_{11}$ and $\sigma_{22}$ represent the normal stresses along *x'*- and *y'*-axes, respectively. The parameters $\sigma_{12}$ and $\sigma_{21}$ represent the shear stresses in the *x'*-*y'* plane.

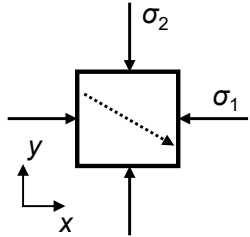

(a) Principal stresses in the original coordinate system.

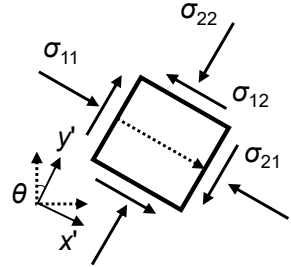

(b) Normal and shear stresses in the rotated coordinate system.

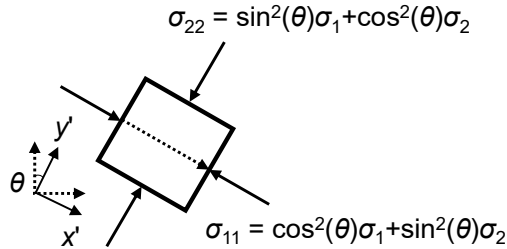

(c) Simplified bi-axial stress state by neglecting the shear stresses in the rotated coordinate system.

Figure 2. Coordinate system transformation (solid arrow: stress direction; dotted arrow: longitudinal wave propagation direction; $\theta$ represents the rotation angle to orient the coordinate system according to the propagation direction).

Based on our previous research [15, 16], it has been established that the velocity change of a longitudinal wave is primarily influenced by the normal stress/strain, while the impact of shear stress/strain on the longitudinal wave velocity can be neglected in the scenario of the plane stress state. As a result, the stress state depicted in Figure 2(b) can be simplified to a bi-axial stress state, as illustrated in Figure 2(c). This figure demonstrates that the acoustoelastic parameter associated with an inclined propagating longitudinal wave closely resembles that in a bi-axial stress state with axial stresses of $\sigma_{11}$ and $\sigma_{22}$. Therefore, along with the normal stresses computed in Equation (9), the acoustoelastic effect in the bi-axial stress state in Figure 2(c) can be formulated as:

$$\rho v^2 = \lambda + 2\mu + (3\lambda + 6\mu + 2l + 4m)\frac{\sigma_{11}}{E} - 2(\lambda + 2l)\frac{\upsilon\sigma_{11}}{E}$$
$$- (3\lambda + 6\mu + 2l + 4m)\frac{\upsilon\sigma_{22}}{E} + (\lambda + 2l)\left(\frac{\sigma_{22}}{E} - \frac{\upsilon\sigma_{22}}{E}\right)$$
$$= \lambda + 2\mu + \frac{3\lambda + 6\mu + 2l + 4m - 2\upsilon(\lambda + 2l)}{E}\left[\cos^2(\theta)\sigma_1 + \sin^2(\theta)\sigma_2\right]$$
$$+ \frac{\lambda + 2l - \upsilon(4\lambda + 6\mu + 4l + 4m)}{E}\left[\sin^2(\theta)\sigma_1 + \cos^2(\theta)\sigma_2\right], \quad (10a)$$

Equation (10a) can also be presented in a form similar to Equation (8) as:

$$v = v^*\sqrt{1 + \left[A_1\cos^2(\theta) + A_2\sin^2(\theta)\right]\sigma_1 + \left[A_1\sin^2(\theta) + A_2\cos^2(\theta)\right]\sigma_2}, \quad (10b)$$

Please note that Equation (10) is not new, and one can find it in the papers dealing with the acoustoelastic effect of waves propagating inclined to the principal stress directions in metals [10-14]. The main assumption in Equation (10) is that the influence of shear stresses on longitudinal wave velocity is negligible, which has not been proven valid in the aforementioned articles. Our previous work [15, 16] theoretically demonstrated the validity of Equation (10). We emphasize that the procedure adopted in this section is not applicable to transverse waves since shear stresses significantly affect transverse wave velocity when bulk waves propagate on the shear deformation plane.

In a special loading scenario where the uniaxial principal stress is along the *y*-axis, Equation (10) can be simplified as:

$$\rho v^2 = \lambda + 2\mu + \frac{3\lambda + 6\mu + 2l + 4m - 2\upsilon(\lambda + 2l)}{E}\sin^2(\theta)\sigma_2$$
$$+ \frac{\lambda + 2l - \upsilon(4\lambda + 6\mu + 4l + 4m)}{E}\cos^2(\theta)\sigma_2. \quad (11)$$

Equation (11) can be expressed in a form similar to Equation (4) and (6) as:

$$v = v^*\sqrt{1 + A_{\text{eff}}\sigma_2}, \quad (12a)$$

where $A_{\text{eff}}$ is the effective acoustoelastic parameter for inclined propagating longitudinal waves:

$$A_{\text{eff}} := \sin^2(\theta)A_1 + \cos^2(\theta)A_2. \quad (12b)$$

As shown in Equation (12), this effective acoustoelastic parameters can be calculated through the relationship between applied stress and longitudinal wave velocity. This loading scenario will be presented in Section 4.

## 4. Experimental validation design
### 4.1 Concrete sample, sensor layout and measurement protocol
The test specimen is a concrete cylinder with dimensions of 300 mm in diameter and 500 mm in height. The cylinder is cast using self-compacting concrete of grade C60/75, and the specific mixture details can be found in Table 3. The mean cube compressive strength (tested from 150 mm cubes) and mean prism elastic modulus of concrete on 28 day are 75 MPa and 39 GPa, respectively.

Table 3. The mixture composition of concrete.

| Component | Amount for 40 liters |
|---|---|
| Sand (0-4 mm) | 33.2 kg |
| Gravel (4-16 mm) | 29.2 kg |
| CEM IIIB | 11.4 kg |
| CEM IIIA | 11.4 kg |
| Fillers (fly ash) | 1.8 kg |
| Water | 7.68 kg |
| Super plasticizer | 172 g |

In the experiment, a total of 11 ultrasonic sensors with a central frequency of 80 kHz were employed as both transmitters and receivers. Specifically, two sensors were affixed to one side of the concrete cylinder as transmitters, while the remaining sensors were placed on the opposite side as receivers, as illustrated in Figure 3. The uniaxial compressive load was applied onto the top and bottom surfaces of the cylindrical sample, aligned with the $y$-axis according to the coordinate system in Figure 3. The transmitter located near the top surface is denoted as Transmitter 1, and the corresponding arrangement of receivers is referred to as SL 1. Likewise, when utilizing the other transmitter, Transmitter 2, the layout of receivers is termed SL 2. The inclinations of the receivers within SL 1 and SL 2 can be found in Table 4.

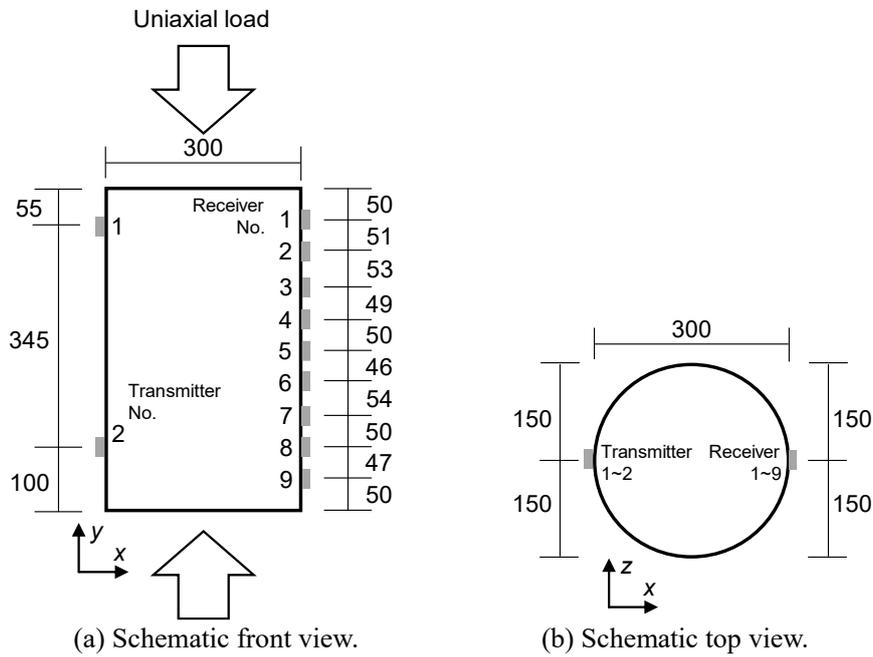

(a) Schematic front view.   (b) Schematic top view.

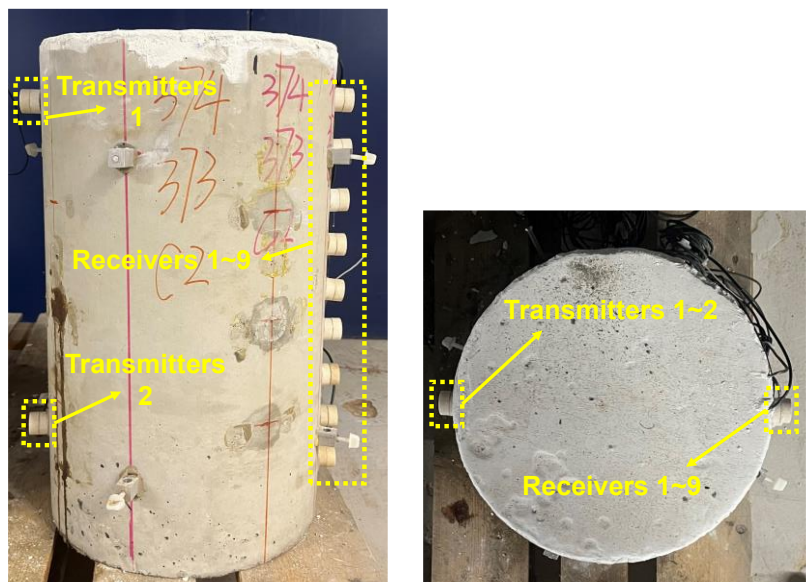

(c) Photos of the specimen and sensors.
Figure 3. Specimen and sensor layout.

Table 4. Receiver inclination for SL1 and SL2 with respect to the *x*-axis on the *x-y* plane (please be aware that 90° indicates that the propagation direction is parallel to the principal stress direction).

|  | Receiver | 1 | 2 | 3 | 4 | 5 | 6 | 7 | 8 | 9 |
|---|---|---|---|---|---|---|---|---|---|---|
| Inclination $\theta$ [°] | SL 1 | 0.96 | 8.72 | 18.26 | 26.26 | 33.43 | 39.12 | 44.81 | 49.24 | 52.78 |
|  | SL 2 | 49.40 | 44.90 | 39.35 | 33.29 | 26.11 | 18.61 | 8.90 | 0.57 | 9.46 |

A total of four repetitive ultrasonic tests, numbered as Test 1–4, were conducted on the concrete sample. The details of each test can be found in Table 5. The age of the concrete during Test 1 is 15 months, while for Test 2, 3 and 4, it is 16 months. All measurements are performed throughout the loading process at a sampling rate of 10 MHz. Prior to the measurement, the sample undergoes three-cycle pre-loading, where the stress ranged from 2.829 MPa to 11.318 MPa, as indicated in Table 5. This pre-loading procedure aims at mitigating the relatively lower slope of the stress-velocity relationship that may occur due to crack formation during the initial loading phase [49]. For the first three tests, the receiver configuration of SL 1 is utilized, with Transmitter 1 serving as the source. Although the minimum and maximum compressive stresses are the same for all three tests, the stress interval in Test 3 is twice as large as that in Test 1 and Test 2. This variation in stress interval is implemented to assess the sensitivity of the obtained results with different stress intervals. The loading protocols of four tests can be found in Figure 4. In Test 4, Transmitter 2 is used as the source, while maintaining the same stress range and interval as Test 3.

Table 5. Detailed information for four tests.

| Test No. | Sensor layout | Minimum compressive stress [MPa] | Maximum compressive stress [MPa] | Stress interval [MPa] | Number of measurements |
|---|---|---|---|---|---|
| 1 | SL 1 | 2.829 | 11.318 | 0.707 | 13 |
| 2 | SL 1 | 2.829 | 11.318 | 0.707 | 13 |
| 3 | SL 1 | 2.829 | 11.318 | 1.415 | 7 |
| 4 | SL 2 | 2.829 | 11.318 | 1.415 | 7 |

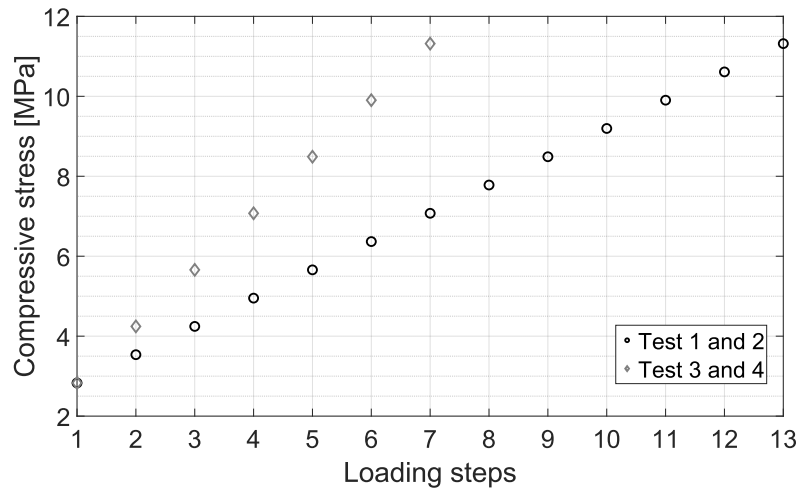

Figure 4. Loading protocol of four tests.

## 4.2 Data processing approach

Since wave interferometry has a better performance than the UPV in detecting stress-induced velocity changes in concrete, the velocity change in this study is obtained using a time-domain-based wave interferometry technique known as the stretching technique [24, 32]. This technique assumes a linear variation of the lag time in a single time window [50] and calculates the velocity change by searching for the stretching factor that maximizes the similarity between the reference signal and the stretched or compressed signal. For a more comprehensive understanding of its theoretical background, we refer to the paper by Lobkis and Weaver [51] as well as our previous work [52]. In this paper, the step-wise stretching technique [53] is employed, where the signal from the previous load step is taken as the reference. The velocity relative to the measurement at the initial load step is then calculated using the following equation:

$$\frac{v_k}{v_1} = \frac{v_2}{v_1}\frac{v_3}{v_2}\cdots\frac{v_k}{v_{k-1}}$$

$$= \left[\frac{v_2-v_1}{v_1}+1\right]\left[\frac{v_3-v_2}{v_2}+1\right]\cdots\left[\frac{v_k-v_{k-1}}{v_{k-1}}+1\right] \quad (13)$$

$$= \prod_{i=1}^{k-1}\left[\left(\frac{dv}{v}\right)_i+1\right],$$

where $v_1$ and $v_k$ indicate the velocity of longitudinal waves propagating in the $x$ direction measured at the first stress level and the $k$th stress level, respectively. The parameter $(dv/v)_i$ can be directly obtained using the stretching technique. Since the interconversion between longitudinal and transverse waves could occur at each scattering event [54], the longitudinal wave component may contain a certain amount of transverse waves. In order to minimize the impact of transverse waves on the analysis, only the initial one-and-a-half distinguishable cycles of the signal will be utilized for the stretching calculation. This corresponds to approximately 23.5 μs in duration at a frequency of 64 kHz.

The effective acoustoelastic parameters calculated using Equation (12) assumes an initial stress of zero. However, during the experiment, the initial compressive stress is 2.82 MPa, as shown in Table 5. Therefore, the following equation, derived from Equation (12), is employed to incorporate the influence of this initial stress:

$$\left(\frac{v_k}{v_1}\right)^2 = \left(\frac{v^*\sqrt{1+A_{\text{eff}}\sigma_{2,k}}}{v^*\sqrt{1+A_{\text{eff}}\sigma_{2,1}}}\right)^2 \quad (14)$$

$$= \frac{A_{\text{eff}}}{1+A_{\text{eff}}\sigma_{2,1}}\sigma_{2,k} + \frac{1}{1+A_{\text{eff}}\sigma_{2,1}},$$

where $\sigma_{2,1}$ and $\sigma_{2,k}$ indicate the principal compressive stress in the $y$ direction at the first stress level (2.82 MPa), and stress level at which the measurement $k$ is taken, respectively. The parameter $v_k$ and $v_1$ represent the wave velocity at measurement $k$ and the wave velocity at the initial stress state, respectively. The effective acoustoelastic parameter can be calculated by means of the slope of compressive stress-square of relative velocity relation, denoted as $s$, through:

$$A_{\text{eff}} = \frac{s}{1-s\sigma_{2,1}}. \quad (15)$$

The slope $s$ can be determined by performing a linear fit of the square of relative velocity with respect to the compressive stress.

## 5. Experimental results of effective acoustoelastic parameters for longitudinal waves in concrete

The signals received from Receivers 1, 5, and 9 during Test 1 under compressive stresses of 2.829 MPa, 5.659 MPa, 8.488 MPa, and 11.318 MPa are shown in Figure 5. In these figures, the waveforms exhibit remarkable similarity across different stress levels, with diminishing amplitudes attributed to increased travel distance. The velocities of ballistic waves in all cases consistently exceed 4000 m/s, indicative of longitudinal waves. As mentioned above, the longitudinal wave component may contain a certain amount of transverse waves because of the interconversion between longitudinal and transverse waves [54]. However, since we apply the stretching technique in a very short time window, around 23.5 μs, after the first arrival, we anticipate minimal influence of transverse waves on the obtained results. While phase shifts in the signals received from Receiver 1 remain insignificant, they become progressively more pronounced with greater inclination, as shown in Figure 5(b) and 5(c). Similar phenomena as described above are also observable in Tests 2, 3 and 4.

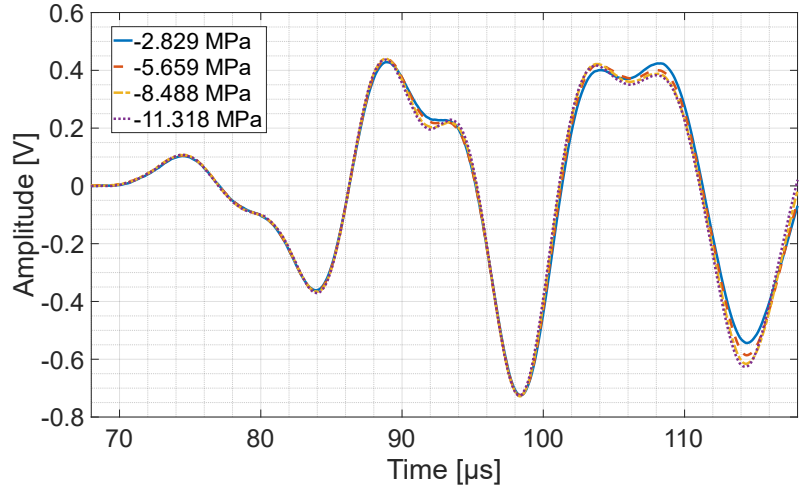

(a) Receiver 1 ($\theta$=0.96°).

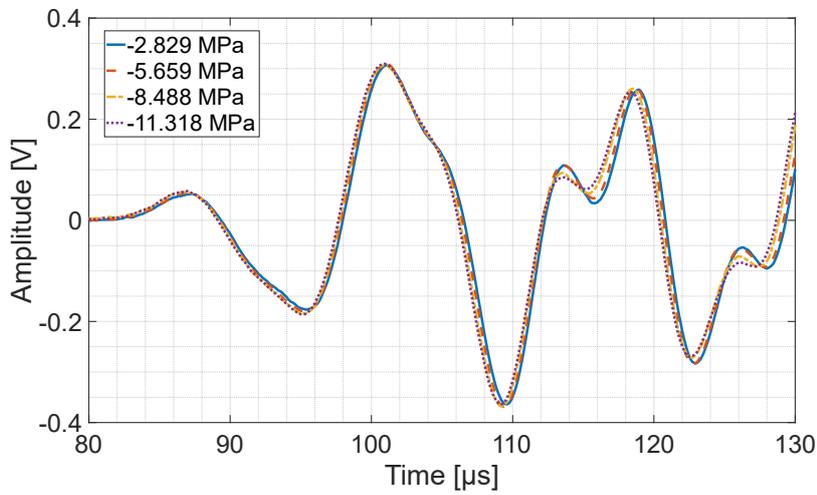

(b) Receiver 5 ($\theta$=33.43°).

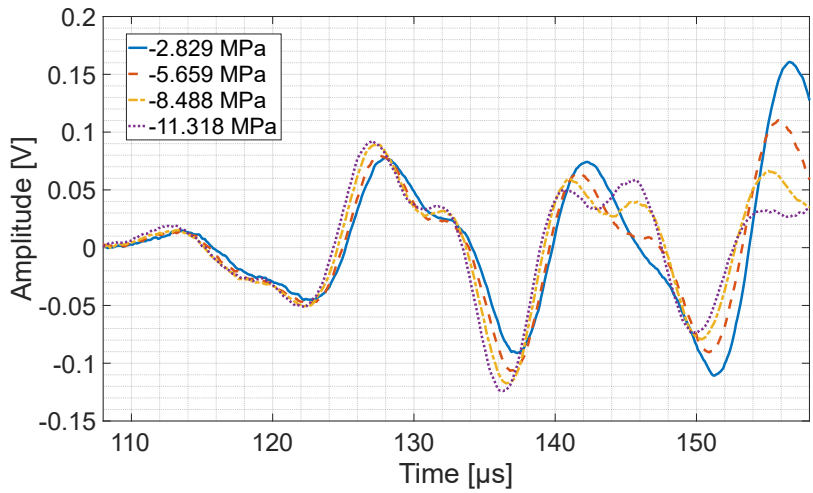

(c) Receiver 9 ($\theta$=52.78°).

Figure 5. Signals from Receiver 1, 5, and 9 during Test 1 under compressive stresses of 2.829 MPa, 5.659 MPa, 8.488 MPa, and 11.318 MPa.

The square of relative velocities as a function of the compressive stress in the four tests are shown in Figure 6. When the inclination approaches 0°, the longitudinal wave propagates nearly perpendicular to the principal stress direction. In this scenario, the compressive stress does not significantly affect longitudinal wave velocity. This finding aligns with observations in the literature [44]. As the inclination increases, a progressively more significant acoustoelastic effect can be observed, characterized by an increasing slope.

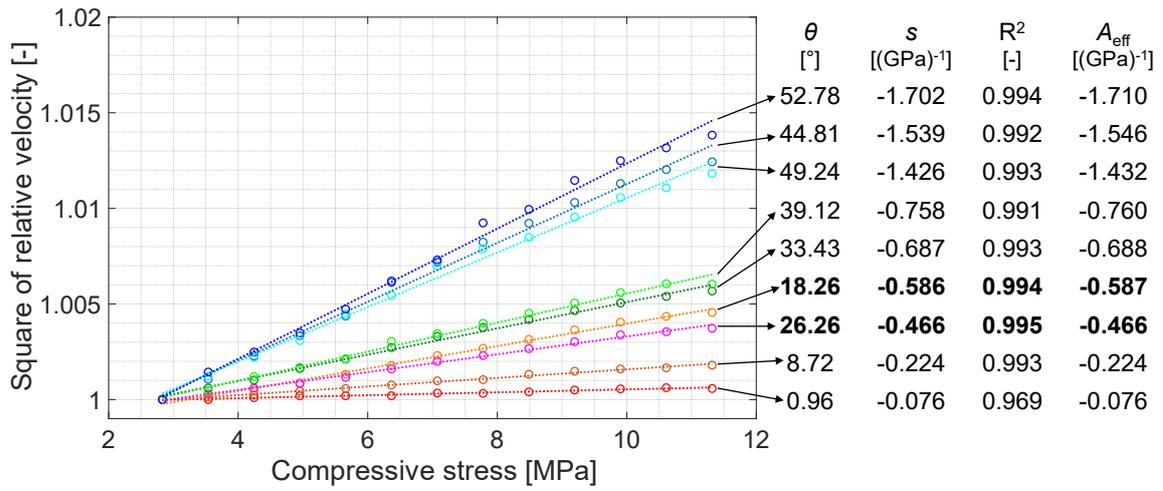

(a) Test 1.

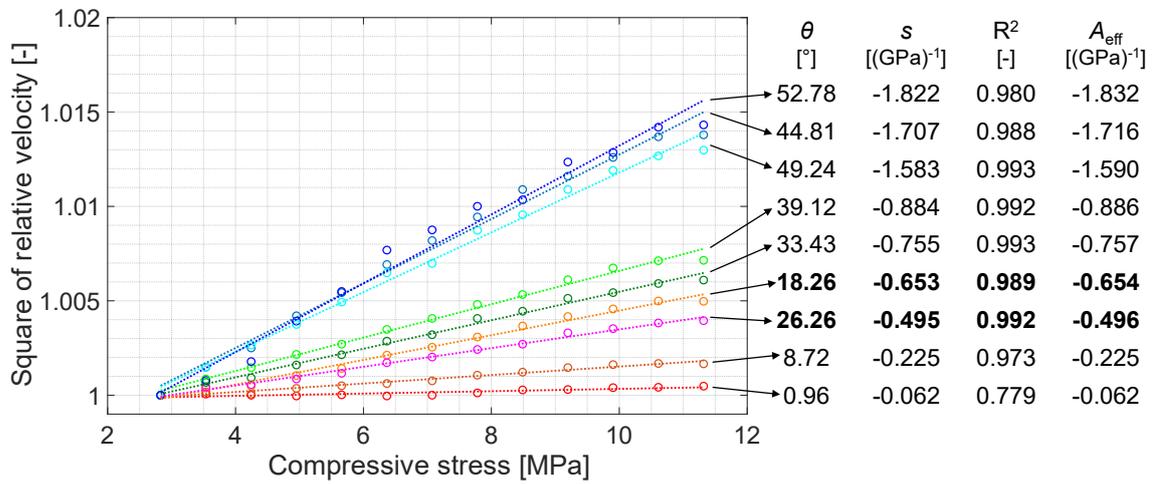

(b) Test 2.

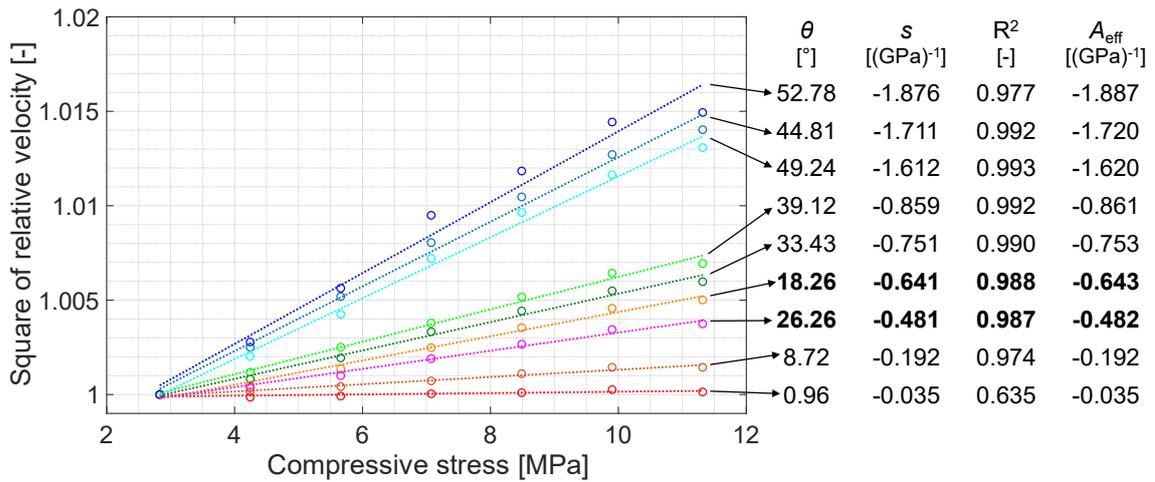

(c) Test 3.

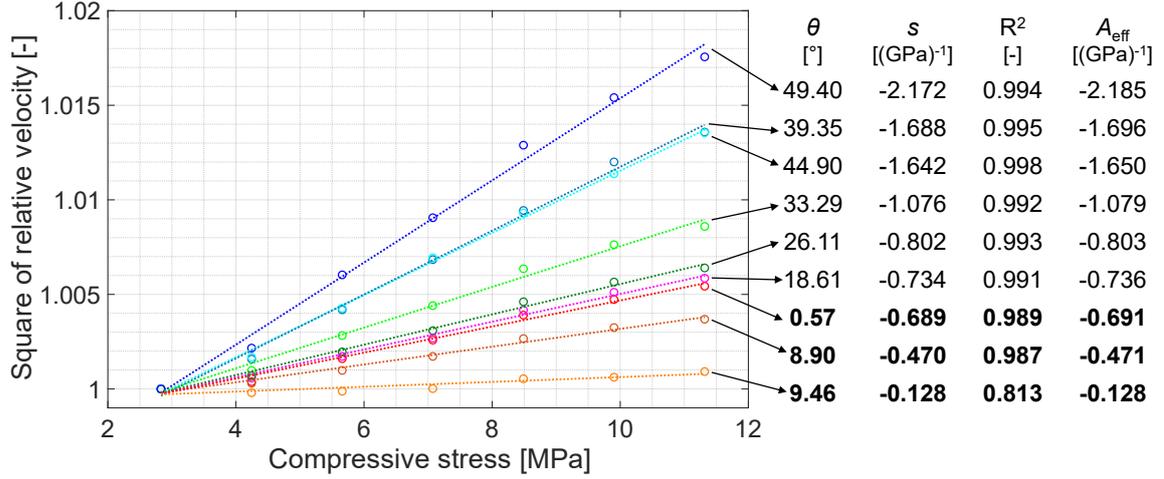

(d) Test 4.

Figure 6. Square of relative velocity vs. compressive stress. The circles represent the experimental measurements, while the dotted lines indicate the linear fit results. The parameter $\theta$ indicates the inclination of sensor-receiver pair with respect to *x*-axis in Figure 3(a). A linear fit of the square of relative velocity with respect to the compressive stress is performed, and the obtained slope is denoted as *s*. The goodness-of-fit is evaluated using the coefficient of determination $R^2$. The effective acoustoelastic parameter $A_{eff}$ is acquired using Equation (15) with $\sigma_{2,1}$ = -2.829 MPa. The bolded results indicate unexpected fluctuations of effective acoustoelastic parameters.

To quantify the relationship between the compressive stress and the square of relative velocity, a linear fit is performed to acquire the slope of this relationship. The goodness-of-fit is evaluated using the coefficient of determination $R^2$ [55, 56]. It is noteworthy that the majority of these coefficients are high, as shown in Figure 6, suggesting a robust linear relationship between compressive stress and the square of relative velocity, which aligns with the theoretical analysis presented in Section 3. This observation further reinforces the accuracy and reliability of the measurements.

The obtained slope is then utilized to calculate the effective acoustoelastic parameter using Equation (15). These parameters can be found in Figure 6. The consistent outcomes and fluctuations observed in the first three tests, demonstrate the reproducibility of the acoustoelastic effect from the same sensor configuration. While the overall trend shows that the magnitudes of $A_{eff}$ increase consistently with increasing inclination, this relationship is reversed for Receiver 3 ($\theta$=18.26°) and Receiver 4 ($\theta$=26.26°). One explanation of the fluctuations in the measurements could be the spatial variation of mechanical properties of concrete [45]. This observation also suggests that relying on a single transmitter-receiver pair to obtain the acoustoelastic parameter may not provide a representative result for the sample. Additionally, examining Test 2 and Test 3 reveals that the effective acoustoelastic parameter is not significantly affected by the magnitude of the load step. Test 4 exhibits slightly different results compared to the initial three tests. Considering that the wave trajectories of transmitter-receiver pairs in Test 4 are different from the previous three tests, this variance could also be ascribed to the spatial variation of mechanical properties of concrete. Nevertheless, the overall trend of the effective acoustoelastic parameters is evident: the magnitude of effective acoustoelastic parameters increases as the inclination increases.

## 6. Validity of acquired effective acoustoelastic parameters

In Section 5, we present the effective acoustoelastic parameters obtained from the experiment for longitudinal waves. To further validate these parameters, we calculate acoustoelastic parameters $A_1$ and $A_2$ based on the results presented in Section 5 and compare them with values reported in the literature.

The effective acoustoelastic parameter is determined by three variables as shown in Equation (12): $A_1$ and $A_2$, which are acoustoelastic parameters, and $\theta$, the inclination of wave propagation. Since the inclination of wave propagation can be inferred from the sensor locations in the experiment, only two acoustoelastic parameters remain unknown. Theoretically, these two parameters can be derived using two effective acoustoelastic parameters corresponding to distinct inclinations of wave propagation. In this study, each test involves longitudinal waves propagated at nine distinct inclinations, resulting in nine effective acoustoelastic parameters. Utilizing these nine parameters alongside the minimum norm least-squares criterion facilitates the determination of the two unknown acoustoelastic parameters, $A_1$ and $A_2$. Once these parameters are obtained, it becomes feasible to reconstruct the effective acoustoelastic parameters for different receivers based on their known inclinations. The deviation between these reconstructed results and the experimental data is evaluated using the root mean square deviation (RMSD). The acoustoelastic parameters, $A_1$ and $A_2$, and RMSD values comparing experimental and reconstructed results are detailed in Table 6. Figure 7 illustrates the comparison between the effective acoustoelastic parameters derived from experiments and those reconstructed using the parameters listed in Table 6.

Table 6. Acoustoelastic parameters obtained using minimum norm least-squares criterion and RMSD values between experimental and reconstructed results.

| Test No. | Acoustoelastic parameters [(GPa)$^{-1}$] | | RMSD values between experimental and reconstructed results [(GPa)$^{-1}$] |
|---|---|---|---|
| | $A_1$ | $A_2$ | |
| 1 | -2.4838 | -0.1155 | 0.1698 |
| 2 | -2.7425 | -0.1192 | 0.1804 |
| 3 | -2.8100 | -0.0859 | 0.1865 |
| 4 | -3.2680 | -0.3629 | 0.1909 |

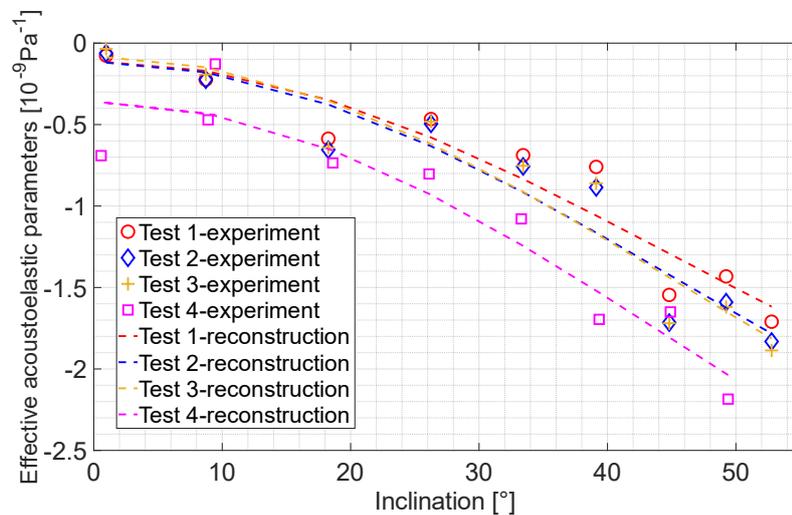

Figure 7. Comparison of effective acoustoelastic parameters obtained from experiments and those reconstructed using the acoustoelastic parameters listed in Table 6.

In the first three tests, which share the same sensor configuration (SL 1), the reconstructed results are quite consistent. Additionally, the reconstructed effective acoustoelastic parameters in Test 4 consistently manifest a higher magnitude compared to those in Tests 1, 2, and 3. As previously mentioned, this discrepancy could be attributed to the spatial variation of mechanical properties in concrete. The acoustoelastic parameter $A_2$ exhibits a significantly lower magnitude than $A_1$ in all four tests, and the magnitude of $A_2$ obtained from Test 4 is at around three times larger than those from the other tests. Despite some fluctuations, the overall trend of the effective acoustoelastic parameters derived from the experiment aligns with the expectation: when the propagation direction of longitudinal waves approaches closer to the direction of the principal stress, the acoustoelastic effect becomes more significant, and the magnitude of the effective acoustoelastic parameters increases.

The acoustoelastic parameters obtained in this study are further compared with those reported in the literature. To ensure comparability, the slope of the stress-velocity change relationship, which is widely adopted in the literature, is converted into the acoustoelastic parameters $A_1$ and $A_2$. The following equation, which is derived from Equation (4), is adopted for longitudinal waves propagating parallel to the uniaxial principal stress direction:

$$\frac{d\left(\frac{v-v^*}{v^*}\right)}{d\sigma_1} = \frac{A_1}{2}\frac{1}{\sqrt{1+A_1\sigma_1}} \quad (16)$$

$$\approx \frac{A_1}{2}.$$

Please note that the sign of approximate equivalence holds as the magnitude of $A_1\sigma_1$, approximately 0.002, is much smaller than 1 in concrete [42]. A comparable expression to Equation (16) can be employed to convert the slope of the stress-velocity change relationship for longitudinal waves propagating perpendicular to the uniaxial principal stress direction into $A_2$, although it will not be presented here. The comparison between the acoustoelastic parameters in this study and those reported in the literature is presented in Table 7. The acoustoelastic parameters in the four tests are consistent with the findings reported by other researchers in the literature, reinforcing the validity of the effective acoustoelastic parameters shown in Section 5.

Table 7. Comparison between the acoustoelastic parameters in this paper and those reported in the literature.

| | | Test 1 | Test 2 | Test 3 | Test 4 | Lillamand et al. [44] | Nogueira et al. [42] | Zhong et al. [33] |
|---|---|---|---|---|---|---|---|---|
| Acoustoelastic parameters [(GPa)$^{-1}$] | $A_1$ | -2.48 | -2.74 | -2.81 | -3.27 | -2.60 | Vary from -0.80 to -5.26 | -2.42 |
| | $A_2$ | -0.12 | -0.12 | -0.09 | -0.36 | -0.40 | Vary from +0.48 to -0.74 | -0.42 |

## 7. Discussion
### 7.1 Error estimation
The precision of the acoustoelastic parameters of concrete obtained through the approach proposed in this paper is influenced by four main factors: (1) simplification of the acoustoelastic theory, (2) accuracy of the stretching technique to estimate velocity changes in concrete, (3) change of mechanical properties of concrete relating to the age of concrete, and (4) the spatial variation of mechanical properties of concrete. The first factor stems from neglecting the impact of shear stresses on longitudinal wave velocities. However, this discrepancy in velocity change is very limited, typically less than 0.06‰ per 1 MPa at a 45° inclination (the maximum observed difference) in the plane stress state, based on previous observations [15, 16].

The second type of error arises from the stretching technique. Based on studies by Weaver et al. [57] and Mao et al. [58], the error magnitude ranges from 0.01‰ to 0.2‰, similar to the magnitude of velocity change when the longitudinal wave propagates perpendicular to the uniaxial stress direction.

To provide a more intuitive understanding of how errors stemming from the stretching technique affect the acquired effective acoustoelastic parameters, we conducted a preliminary error analysis. We assume a maximum error of 0.1‰ in $dv/v$, a value consistent with findings by Weaver et al. [57] and Mao et al. [58]. Using Taylor expansion, we estimate the error of $(dv/v+1)^2$ to be approximately 0.2‰. In this demonstration, we use Test 1 as an illustration. The acoustoelastic parameter $A_2$ characterizes velocity changes in the medium when longitudinal waves propagate perpendicular to the stress direction. In Test 1, this propagation direction closely aligns with the sensor inclination of 0.96° (Receiver 1). The relationship between the square of relative velocity and compressive stress for this sensor inclination, along with the errors resulting from the stretching technique, can be seen in Figure 8(a). The acoustoelastic parameter $A_1$ describes velocity changes in the medium when longitudinal waves travel parallel to the stress direction. However, this paper lacks a sensor inclined in such a manner. Instead, for illustrative purposes, we utilize the sensor inclination of 52.78° (Receiver 9), which is closest to the scenario of $A_1$ in our tests. The relationship between the square of relative velocity and compressive stress for this sensor inclination, together with the stretching technique-induce error, is displayed in Figure 8(b).

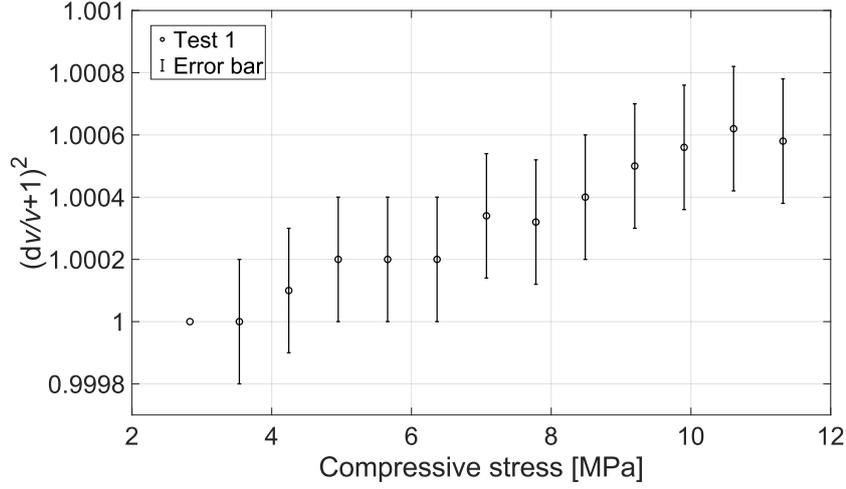

(a) Inclination of receivers of 0.96° (Receiver 1).

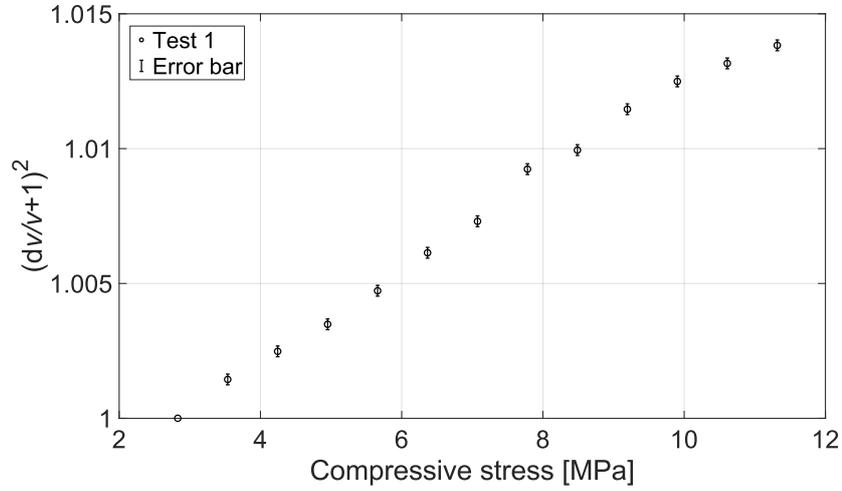

(b) Inclination of receivers of 52.78° (Receiver 9).

Figure 8. Square of relative velocity vs. compressive stress in Test 1 with receivers inclined at 0.96° (Receiver 1) and 52.78° (Receiver 9). The center circle of the error bar indicates the square of relative velocity acquired from Test 1. The absolute maximum error of $dv/v$ is 0.1‰.

Figure 8 clearly illustrates that errors propagated from the stretching technique significantly affect measurements from Receiver 1 but have minimal impact on those from Receiver 9. To further quantify the propagation of these errors to the effective acoustoelastic parameter, the following equation will be employed to estimate the errors in the effective acoustoelastic parameter, denoted as $\Delta A_{\text{eff}}$:

$$\Delta A_{\text{eff}} = \frac{\partial}{\partial s}\left(\frac{s}{1-s\sigma_{2,1}}\right)\Delta s$$
$$= \frac{\Delta s}{\left(1-s\sigma_{2,1}\right)^2}, \quad (17)$$

where $\Delta s$ indicates the error in the fitted slope, propagated from the errors in the velocity change acquired using the stretching technique. The slope $s$ can be determined by conducting a linear fit of the square of relative velocity to the compressive stress. The error in the fitted slope $\Delta s$ can be estimated using the following equation:

$$\Delta s \approx \left| \frac{(y_2 + \Delta) - (y_1 - \Delta)}{x_2 - x_1} - \frac{y_2 - y_1}{x_2 - x_1} \right|$$
$$= \left| \frac{2\Delta}{x_2 - x_1} \right|, \qquad (18)$$

where $x_1$ and $x_2$ represent two query points, while $y_1$ and $y_2$ denote corresponding fitted values at these query points. The parameter $\Delta$ indicates the absolute error associated with the two fitted values. In this study, $\Delta$ is equal to the error of $(dv/v+1)^2$, approximately 0.2‰. The parameters $x_1$ and $x_2$ correspond to compressive stresses at the second stress level, 3.54 MPa, and the last stress level, 11.32 MPa, respectively. Utilizing Equation (17) and (18), the propagated error from the stretching technique to the effective acoustoelastic parameter is calculated. When the inclination of receivers is 0.96° (Receiver 1), the propagated error is determined as 0.051 GPa. This value is comparable in magnitude to the effective acoustoelastic parameter acquired from the same receiver in Test 1, -0.076 GPa. The proportion of the magnitude of error to the magnitude of effective acoustoelastic parameter is approximately 67.1%. Furthermore, the effective acoustoelastic parameters of Receiver 1 obtained from Test 2 (-0.062 GPa) and Test 3 (-0.035 GPa), utilizing identical transmitter configuration, fall within the specified error bounds. In contrast, when the inclination of receivers is 52.78° (Receiver 9), the propagated error to the effective acoustoelastic parameter is computed to be 0.052 GPa. Notably, this magnitude is significantly smaller than the effective acoustoelastic parameter acquired from the same receiver in Test 1, -1.710 GPa. The proportion of the magnitude of error to the magnitude of effective acoustoelastic parameter is only 3.0%. However, utilizing effective acoustoelastic parameters obtained from receivers with lower inclinations does not substantially affect the values of acoustoelastic parameters $A_1$ and $A_2$. Table 8 displays the acoustoelastic parameters derived using effective parameters from receivers with inclinations larger than 10°. These parameters exhibit similar magnitudes compared to those in Table 6, indicating that the smaller effective parameters from receivers with lower inclinations minimally influence the fitting process based on the minimum norm least-squares criterion.

Table 8. Acoustoelastic parameters obtained using effective acoustoelastic parameters acquired by receivers with inclinations larger than 10°.

| Test No. | Receivers involved | Acoustoelastic parameters [(GPa)$^{-1}$] | |
|---|---|---|---|
| | | $A_1$ | $A_2$ |
| 1 | Receiver 3~9 | -2.4933 | -0.1062 |
| 2 | Receiver 3~9 | -2.7289 | -0.1307 |
| 3 | Receiver 3~9 | -2.8008 | -0.0934 |
| 4 | Receiver 1~6 | -3.2288 | -0.3720 |

Furthermore, the parameters of Receiver 9 in Test 2 (-1.832 GPa) and Test 3 (-1.887 GPa) surpass the error bounds, which, based on the discussion above, range from -1.710±0.052 GPa. There is a notable increase in the magnitude of effective acoustoelastic parameters from Test 1 to Test 2 (0.122 GPa), whereas from Test 2 to Test 3, this difference is lower (0.055 GPa). Interestingly, the parameter change between Test 2 and Test 3 slightly exceeds the estimated propagated error, suggesting that the actual error from the stretching technique might be higher than 0.1‰ in concrete. Notably, there is no existing research on the precision of the stretching technique when applied to concrete. Two potential reasons behind the significant jump in the magnitude of effective acoustoelastic parameters acquired from Receiver 9 between Test 1 and Test 2 include the underestimated error while using the stretching technique, as discussed above, and the temperature fluctuation, especially considering that the almost identical effective acoustoelastic parameters for Test 2 and Test 3. Despite these considerations, it can be concluded that errors stemming from the stretching technique notably affect acoustoelastic parameters when the longitudinal wave propagation direction is perpendicular or nearly perpendicular to the uniaxial stress direction, while their impact is limited when the longitudinal wave propagates parallel or nearly parallel to the uniaxial stress direction.

The third aspect to consider is the age of the concrete. The mechanical properties of concrete, which can be represented by second-order elastic constants Lamé parameters, will gradually develop with time, and such a development is called hydration. The increase of Lamé parameters during the hydration process results in an increase of both longitudinal and transverse wave velocities [59]. However, there is no research on the time-dependent behaviour of Murnaghan constants in concrete, but it is certain that these constants must change during hydration. Understanding the evolution of Lamé parameters and Murnaghan constants during concrete hydration could aid in interpreting the notable increase in the magnitude of effective acoustoelastic parameters between Test 1 and Test 2.

The last aspect concerns the spatial variation of mechanical properties of concrete [45]. Our study reveals variations in effective acoustoelastic parameters with changes in transmitter position, suggesting differences in parameters across various transmitter and receiver locations, even within the same concrete sample. Importantly, these differences can be substantial.

To enhance the future application of the proposed approach for concrete stress monitoring, two key issues should be addressed: (1) analyzing the errors in the stretching technique for retrieving stress-induced velocity changes in concrete to establish its minimum resolution in retrieving velocity change, and (2) exploring the impact of spatial variation of mechanical properties of concrete on acoustoelastic parameters through numerous repetitive experiments.

### 7.2 Potential applications of the proposed theory
There are two potential applications of the proposed theoretical framework for monitoring concrete infrastructures:
1. Determination of the acoustoelastic parameters for longitudinal waves.
2. Monitoring magnitudes and directions of principal stresses in the plane stress state.

The first application involves using effective acoustoelastic parameters to determine the acoustoelastic parameters $A_1$ and $A_2$, as demonstrated in Section 6. Compared to parameters obtained using a single transmitter-receiver pair, the setup shown in Figure 3(a) offers the advantage of accounting for the spatial variation of mechanical properties, thereby providing more representative acoustoelastic parameters for a given area.

The second application builds on the first one. With $A_1$ and $A_2$ determined, it becomes possible to estimate changes in the magnitudes and directions of principal stresses within concrete structures in the plane stress state. The acoustoelastic expression for longitudinal waves in a biaxial stress condition is presented in Equation (10b). The velocity change of longitudinal waves propagating at an angle to principal stress directions while the biaxial stresses are applied in the $y$- and $x$-axis is then:

$$\frac{v_2 - v_1}{v_1} = \frac{\sqrt{1 + \left[\cos^2(\theta_2)A_1 + \sin^2(\theta_2)A_2\right]\sigma_{1,2} + \left[\sin^2(\theta_2)A_1 + \cos^2(\theta_2)A_2\right]\sigma_{2,2}}}{\sqrt{1 + \left[\cos^2(\theta_1)A_1 + \sin^2(\theta_1)A_2\right]\sigma_{1,1} + \left[\sin^2(\theta_1)A_1 + \cos^2(\theta_1)A_2\right]\sigma_{2,1}}} - 1 , \qquad (19)$$

where $\sigma_{i,1}$ and $\sigma_{i,2}$ indicate the principal stresses in the $i$ direction at the first stress level, and the second stress level, respectively; $\theta_1$ and $\theta_2$ are the inclination of the wave propagation, as shown in Figure 2, at the first stress level, and the second stress level, respectively. Following the calibration to obtain the acoustoelastic parameters $A_1$ and $A_2$, Equation (19) involves six unknowns: $\theta_1$, $\theta_2$, $\sigma_{1,1}$, $\sigma_{1,2}$, $\sigma_{2,1}$, and $\sigma_{2,2}$. Theoretically, a system of six equations is necessary to determine all unknowns, achievable by adjusting sensor configurations in various inclinations. In practice, one potential approach to simplifying Equation (19) is to predetermine the directions and magnitudes of the biaxial principal stresses at the initial stress level. For example, the first measurement can be conducted in a stress scenario that no external load applied, in which the magnitudes of principal stresses are approximately zero. Then the unknowns can be reduced to $\theta_2$, $\sigma_{1,2}$, and $\sigma_{2,2}$. It is important to note that (i) the principal stresses estimated through the ballistic wave portion represent mean principal stresses along the ballistic wave trajectory, and (ii) Equation (19) is solely applicable to the elastic stage of concrete, where no visible crack exists in the medium under examination. The further use of the proposed theory in monitoring magnitudes and directions of principal stresses in concrete structures will be presented in our future work.

## 8. Conclusion
This paper presents a comprehensive study on the acoustoelastic effect of longitudinal waves propagating inclined to the principal stress directions in concrete, integrating both theoretical and experimental investigations. Building upon our previous findings, which demonstrated the negligible influence of shear stresses on longitudinal wave velocity, we propose a simplified acoustoelastic expression specifically tailored for longitudinal waves in a plane stress state. Our theoretical analysis reveals that the effective acoustoelastic parameter of a longitudinal wave propagating inclined to the principal stress directions can be expressed as a combination of the acoustoelastic parameters of waves propagating parallel and perpendicular to the uniaxial stress direction. To validate the proposed acoustoelastic expression, we conducted a uniaxial compression test on a concrete cylinder. Although some experimental observations exhibit fluctuations, the overall trend of effective acoustoelastic parameters aligned with the theory. Additionally, the magnitudes of acoustoelastic parameters back-calculated from effective parameters using the minimum norm least-squares criterion were consistent with those reported in the literature, further supporting the validity of the proposed theory.

**Appendix A. Acoustoelastic parameters in the natural frame.**

The derivation of the equation of motion in the natural frame is shown in this appendix. This derivation is based on our previous derivation [15, 16]. The derivation in this appendix will not start from scratch. Please refer to our previous derivation for more details, i.e., notations.

The equation of motion of acoustoelasticity is defined as:

$$D_{\xi\beta k\delta} \frac{\partial^2 u_k^{\text{dynamic}}}{\partial a_\beta \partial a_\delta} = \rho \frac{\partial^2 u_\xi^{\text{dynamic}}}{\partial t^2} \quad , \tag{A1}$$

where $\rho$ are the mass density in the natural state without load applied, and $u_i^{\text{dynamic}}$ denotes the displacements of dynamic disturbance. The acoustoelastic modulus $D_{\xi\beta k\delta}$ is defined as:

$$D_{\xi\beta k\delta} := C_{\beta\delta\varepsilon\eta} e_{\varepsilon\eta} \delta_{k\xi} + C_{\xi\beta k\delta} + C_{\xi\beta k\delta\varepsilon\eta} e_{\varepsilon\eta} + C_{\alpha\beta k\delta} \frac{\partial u_\xi}{\partial a_\alpha} + C_{\xi\beta\gamma\delta} \frac{\partial u_k}{\partial a_\gamma} \quad . \tag{A2}$$

where $e_{\varepsilon\eta}$ indicates the static strain tensor caused by the external load, and $u_k$ denotes the displacements associated with the static deformation induced by the external load. The $C_{ijkl}$ in Equation (A2) represents the second-order elastic coefficients composed of the second-order elastic constants, also known as Lamé parameters, $\lambda$ and $\mu$. The second-order elastic coefficients can be expressed using Voigt notation as $C_{IJ}$. The third-order elastic coefficients, $C_{\alpha\beta\gamma\delta\varepsilon\eta}$, can be expressed using the third-order elastic constants, i.e., Murnaghan constants $l$, $m$, and $n$. Similar to $C_{ijkl}$, the third-order elastic coefficients $C_{\alpha\beta\gamma\delta\varepsilon\eta}$ can also be represented using Voigt notation as $C_{KLM}$, which are commonly known as standard third-order elastic coefficients. Equation (A1) and (A2) can also be found in [5]. The linear part of the strain will be used to simplify Equation (A2):

$$e_{\alpha\beta} = \frac{1}{2}\left(\frac{\partial u_\beta}{\partial a_\alpha} + \frac{\partial u_\alpha}{\partial a_\beta}\right) \quad . \tag{A3}$$

Additionally, the second- and third-order elastic coefficients will be represented using Voigt notation. Here are the derivation details of acoustoelastic moduli $D_{\xi\beta k\delta}$:

$$\begin{aligned} D_{1111} &= C_{11\varepsilon\eta} e_{\varepsilon\eta} \delta_{11} + C_{1111} + C_{1111\varepsilon\eta} e_{\varepsilon\eta} + C_{\alpha 111} \frac{\partial u_1}{\partial a_\alpha} + C_{11\gamma 1} \frac{\partial u_1}{\partial a_\gamma} \\ &= C_{1111} + C_{1111} e_{11} + C_{1122} e_{22} + C_{1133} e_{33} + C_{111111} e_{11} + C_{111122} e_{22} + C_{111133} e_{33} \\ &\quad + C_{1111} \frac{\partial u_1}{\partial a_1} + C_{1111} \frac{\partial u_1}{\partial a_1} \\ &= C_{11} + (3C_{11} + C_{111}) e_{11} + (C_{12} + C_{112}) e_{22} + (C_{13} + C_{113}) e_{33} \quad . \end{aligned} \tag{A4}$$

$$\begin{aligned} D_{2121} &= C_{11\varepsilon\eta} e_{\varepsilon\eta} \delta_{22} + C_{2121} + C_{2121\varepsilon\eta} e_{\varepsilon\eta} + C_{\alpha 121} \frac{\partial u_2}{\partial a_\alpha} + C_{21\gamma 1} \frac{\partial u_2}{\partial a_\gamma} \\ &= C_{2121} + C_{1111} e_{11} + C_{1122} e_{22} + C_{1133} e_{33} + C_{212111} e_{11} + C_{212122} e_{22} + C_{212133} e_{33} \\ &\quad + C_{2121} \frac{\partial u_2}{\partial a_2} + C_{2121} \frac{\partial u_2}{\partial a_2} \\ &= C_{66} + (C_{11} + C_{661}) e_{11} + (C_{12} + 2C_{66} + C_{662}) e_{22} + (C_{13} + C_{663}) e_{33} \quad . \end{aligned} \tag{A5}$$

$$D_{3131} = C_{11\varepsilon\eta}e_{\varepsilon\eta}\delta_{33} + C_{3131} + C_{3131\varepsilon\eta}e_{\varepsilon\eta} + C_{\alpha131}\frac{\partial u_3}{\partial a_\alpha} + C_{31\gamma1}\frac{\partial u_3}{\partial a_\gamma}$$

$$= C_{3131} + C_{1111}e_{11} + C_{1122}e_{22} + C_{1133}e_{33} + C_{313111}e_{11} + C_{313122}e_{22} + C_{313133}e_{33} \quad (A6)$$

$$+ C_{3131}\frac{\partial u_3}{\partial a_3} + C_{3131}\frac{\partial u_3}{\partial a_3}$$

$$= C_{55} + (C_{11} + C_{551})e_{11} + (C_{12} + C_{552})e_{22} + (C_{13} + 2C_{55} + C_{553})e_{33} \ .$$

$$D_{1121} = C_{11\varepsilon\eta}e_{\varepsilon\eta}\delta_{12} + C_{1121} + C_{1121\varepsilon\eta}e_{\varepsilon\eta} + C_{\alpha121}\frac{\partial u_1}{\partial a_\alpha} + C_{11\gamma1}\frac{\partial u_2}{\partial a_\gamma}$$

$$= C_{112112}e_{12} + C_{112121}e_{21} + C_{2121}\frac{\partial u_1}{\partial a_2} + C_{1111}\frac{\partial u_2}{\partial a_1} \quad (A7)$$

$$= 2C_{166}e_{12} + C_{66}\frac{\partial u_1}{\partial a_2} + C_{11}\frac{\partial u_2}{\partial a_1} = D_{2111} \ .$$

$$D_{1131} = C_{11\varepsilon\eta}e_{\varepsilon\eta}\delta_{13} + C_{1131} + C_{1131\varepsilon\eta}e_{\varepsilon\eta} + C_{\alpha131}\frac{\partial u_1}{\partial a_\alpha} + C_{11\gamma1}\frac{\partial u_3}{\partial a_\gamma}$$

$$= C_{113113}e_{13} + C_{113131}e_{31} + C_{3131}\frac{\partial u_1}{\partial a_3} + C_{1111}\frac{\partial u_3}{\partial a_1} \quad (A8)$$

$$= 2C_{155}e_{13} + C_{55}\frac{\partial u_1}{\partial a_3} + C_{11}\frac{\partial u_3}{\partial a_1} = D_{3111} \ .$$

$$D_{2131} = C_{11\varepsilon\eta}e_{\varepsilon\eta}\delta_{23} + C_{2131} + C_{2131\varepsilon\eta}e_{\varepsilon\eta} + C_{m131}\frac{\partial u_2}{\partial a_m} + C_{21m1}\frac{\partial u_3}{\partial a_m}$$

$$= C_{213123}e_{23} + C_{213132}e_{32} + C_{3131}\frac{\partial u_2}{\partial a_3} + C_{2121}\frac{\partial u_3}{\partial a_2} \quad (A9)$$

$$= 2(C_{654} + C_{44})e_{23} = D_{3121} \ .$$